\documentstyle[aps,twocolumn,prl,epsf]{revtex}
\title{Quantifying Entanglement}
\author{V. Vedral, M.B. Plenio, M.A. Rippin, P. L. Knight}
\address{Optics Section, Blackett Laboratory, Imperial College London, 
London SW7 2BZ, England}
\date{\today}

\begin{document}

\maketitle

\begin{abstract}

We present conditions every measure of entanglement 
has to satisfy and construct a whole class of 'good' entanglement measures. 
The generalization of our class of entanglement measures to more than two
particles is straightforward. We present a measure which has a statistical 
operational basis that might enable experimental determination of the 
quantitative degree of entanglement.
\end{abstract}

We have witnessed great advances in quantum information 
theory in recent years. There are two distinct directions in which progress is 
currently being made: quantum computation and error correction on the 
one hand (for a short survey see \cite{Barenco,Plenio}) and
non--locality, Bell's inequalities and purification on the other hand
\cite{Gisin,Deutsch}. There has also been a number of papers relating the 
two (e.g. \cite{Bennett1,Vedral2}). Our present work belongs to this second 
group. Recently it was realised that the CHSH (Clauser-Horne-Shimony-Holt) form 
of Bell's inequalities 
are not a sufficiently good measure of quantum correlations in the sense that
there are states which do not violate the CHSH inequality, but on the 
other hand can be purified by local interactions and classical communications
to yield a state that does violate the CHSH
inequality \cite{Gisin}. Subsequently, it was shown that the only
states of two two--level systems which cannot be purified are those that 
can be written 
as the sum over density operators which are direct products states of the 
two subsystems \cite{Horodecki1}. Therefore, although it is possible to say 
whether a quantum state is
entangled or not, the amount of entanglement cannot 
easily be determined for general mixed states. Bennett et al \cite{Bennett1}
have recently proposed a measure of entanglement for a general mixed state 
of two quantum subsystems. However, this measure has the disadvantage that it
is hard to compute for a general state, even numerically. 
In this letter we specify conditions which any measure of entanglement has to 
satisfy and construct a whole class of `good' entanglement measures.
Our measures are geometrically intuitive.

Unless stated otherwise the following considerations apply to a system composed 
of two quantum subsystem of arbitrary dimensions.
First we define the term {\em purification procedure} more precisely.   
There are three distinct ingredients in any protocol that aims at increasing 
correlations between two quantum subsystems locally:\\
 
{\em Local general measurements} (LGM): these are performed by the two 
parties ($A$ and $B$) separately and are described by two sets of operators 
satisfying the completeness relations 
$\sum_i {A}^{\dagger}_i {A}_i = \bf{I}$ and $\sum_j {B}^{\dagger}_j {B}_j = 
\bf{I}$. The joint action of the two is described by 
$\sum_{ij}  A_i\otimes  B_j$, which again describes a local general
measurement. 

{\em Classical communication} (CC): this means that the actions of 
$A$ and $B$ can be classically correlated. This can be described by a 
complete measurement on the whole space $A+B$ which, as opposed to local 
general measurements, is not necessarily decomposable into a direct product 
of two operators as above, each acting on only one subsystem. 
If ${\rho}_{AB}$ is the joint state of subsystems $A$ and $B$ then the
transformation involving `LGM+CC' would look like

\begin{equation}
 \rho_{AB} \longrightarrow   \sum_i  A_i \otimes  B_i \; 
{\rho}_{AB} \;  A^{\dagger}_i \otimes  B^{\dagger}_i
\label{1}
\end{equation}
i.e. the actions of $A$ and $B$ are `correlated'. The mapping given in
eq. (\ref{1}) is completely positive. To ensure that it is also trace 
preserving we have to require  
$\sum_{i} A^{\dagger}_i  A_i \otimes  B^{\dagger}_i  B_i=\bf{I}$.
Both LGM and CC are linear transformations on the set
of states. Note that as the third ingredient all purification schemes use 
LGM and CC but also reject part of the original ensemble making the whole
transformation non--linear \cite{Deutsch}. 

We note that all entangled (inseparable)
states can be purified to an ensemble of maximally entangled states 
\cite{Horodecki1}.
This implies that any `good' measure of entanglement has to be zero
if and only if  the state is disentangled (defined by a convex sum of the 
form $\sum_i p_i  \rho_A^i \otimes  \rho_B^i$). Here we would like to 
quantify the degree of entanglement. In the following we briefly review 
some measures of entanglement between two quantum systems (for a review 
of correlation measures see \cite{Ekert2}). 

{\em Entanglement of Creation:}
Bennett et al \cite{Bennett1} define the entanglement of creation of a state 
${\rho}$ by 
\begin{equation}
E( \rho):= \mbox{min} \sum_i p_i S( \rho_A^i)
\end{equation}
where $S( \rho_A)$ is the von Neumann entropy (to be defined in eq. (\ref{3}) 
and the minimum is taken over all the possible realisations of the state, 
$ \rho_{AB} = \sum_j p_j |{\psi_j}\rangle\langle{\psi_j}|$ with $ \rho_A^i =
\mbox{tr}_B (|{\psi_i}\rangle\langle{\psi_i}|)$. The
entanglement of creation cannot be increased by the 
combined action of LGM+CC \cite{Bennett1}. 
 
{\em Entanglement of Distillation }\cite{Bennett1}: This is the number of 
maximally
entangled pairs that can be purified from a given state. This measure 
depends on the particular process of purification and it is not yet clear 
how to compute it in an efficient and unique way.

It seems to be difficult to calculate the degree of entanglement 
for a general state using these two definitions and a closed form 
would be very much desired for further progress \cite{Bennett2}.
The problem is quite involved as one has to minimize over all 
possible decompositions of the density operator in question or over 
all possible purification schemes. There 
are other measures of entanglement which are simpler to calculate but which 
cannot distinguish between quantum and classical correlations. We
discuss two and show how they can be generalised to give 
`good' measures of entanglement; in fact, we show how to derive a whole
class of measures of entanglement.

 {\em Von Neumann Entropy:}
Given a pure state $ \rho_{AB}$ of two subsystems 
A and B we define the states $ \rho_{A}=tr_B\{ \rho_{AB}\}$ and
$ \rho_{B}=tr_A\{ \rho_{AB} \}$ where the partial trace has been taken over 
one subsystem, either A or B. Then the von Neumann entropy of the reduced
density operators is given by
\begin{equation}
	S({\rho}_A) := -tr( {\rho}_A \ln {\rho}_A ) = 
	-tr( {\rho}_B \ln {\rho}_B )\;\; .
	\label{3}
\end{equation}
In the case of a disentangled pure joint state $S({\rho}_A)$ is zero, 
and for maximally entangled states it gives $\ln 2$. However, for mixed states
$ \rho_{AB}$ this measure fails to distinguish classical and quantum 
mechanical correlations.

{\em Von Neumann Mutual Information:} This is defined by:
\begin{eqnarray}
I_N({\rho}_A:{\rho}_B\, ;{\rho}_{AB})  :=  S({\rho}_A) + S({\rho}_B) - 
S({\rho}_{AB})\;\; ,
\label{def7b}
\end{eqnarray}

which essentially reduces to eq. (\ref{3}) for pure states of the joint system 
$ \rho_{AB}$. It is known that $I_N$ {\bf cannot} increase under local general 
measurement only \cite{Vedral2,Lindblad1}, but {\bf can} increase under LGM+CC 
showing that it cannot properly distinguish between classical and quantum 
mechanical correlations. The von Neumann mutual information can intuitively 
be understood as follows: the mutual information calculates a `distance' 
between 
a given state $ \rho_{AB}$ and {\bf one} of its disentangled counterparts
$ \rho_A \otimes   \rho_B$. The crucial word here is `one', as 
there are many other disentangled states for which we could calculate 
$I_N$, which indicates the failure of this 
measure for general mixed states but also suggests its successful
generalization.

Before we generalize the von Neumann mutual information we present necessary
conditions any measure of entanglement $E({\sigma})$ has to satisfy: 
\begin{enumerate}
\item $E({\sigma})=0$ iff ${\sigma}$ is separable.
\item Local unitary operations leave $E({\sigma})$ invariant, i.e.
$E({\sigma})=E({U}_A\otimes {U}_B{\sigma}
{U}_A^{\dagger}\otimes {U}_B^{\dagger})$.
\item The measure of entanglement $E({\sigma})$ cannot increase under 
LGM+CC given by $\Theta$, ie $E(\Theta{\sigma})\le E({\sigma})$.
\end{enumerate}
The origin of condition 1) is that separable states are known to contain 
no entanglement, i.e. they {\em cannot} be purified by LGM+CC to maximally
entangled states, however, any inseparable state can be purified and therefore
contains some entanglement. The reason for condition 2) is that local unitary 
transformations represent a local change of basis only and  
leave quantum correlations unchanged. The reason for condition 3) is that any 
increase in correlations achieved by LGM+CC should be classical in 
nature and therefore entanglement should not be increased.

In the following we construct a new class of measures that satisfy the 
conditions 1)-3).
Let us consider a set $\cal T$ of all density matrices of two quantum 
subsystems, $A$ and $B$ (see Fig. 1) . Let us further divide $\cal T$ into
\begin{figure}[hbt]
\epsfxsize6.0cm
\centerline{\epsfbox{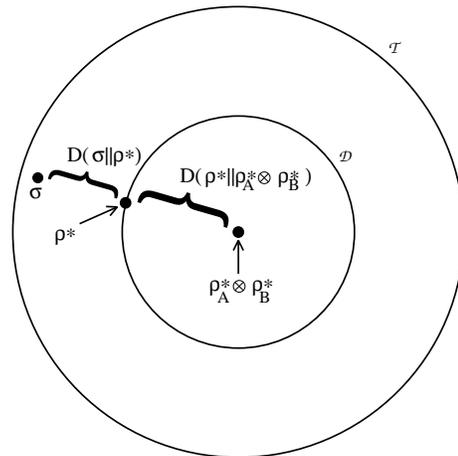}}
\vspace*{0.0cm}
\caption{The set of all density matrices, ${\cal T}$ is represented by the 
outer circle. Its subset, a set of disentangled states ${\cal D}$ is 
represented by the inner circle. A state $ \sigma$ belongs to the 
entangled states, and $ \rho^*$ is the disentangled state that minimizes 
the distance $D( \sigma || \rho)$, thus representing the amount 
of quantum correlations in $\sigma$. State 
$ \rho^*_A \otimes  \rho^*_B$ is obtained by tracing $ \rho^*$ 
over $A$ and $B$. $D( \rho^* || \rho^*_A \otimes  \rho^*_B)$ 
represent the classical part of correlations in the state $ \sigma$. }
\end{figure}
two disjunctive subsets: a set containing
all disentangled states--hereafter labelled by $\cal D$, 
and a set of all the entangled states (all states in $\cal T  - \cal D$) -- 
hereafter labelled by $\cal E$. 
Note that both $\cal T$ and $\cal D$ ({\bf but not} $\cal E$!) are convex 
sets, i.e. $ \rho_1\, ,  \rho_2 \in {\cal T}({\cal D}) \Rightarrow 
\lambda  \rho_1 + (1-\lambda) \rho_2 \in {\cal T}({\cal D})$. The 
entanglement of a matrix 
$ \sigma \in {\cal T}$ will now be defined as:

\begin{equation}
E({\sigma}):= \min_{ \rho \in \cal D}\,\,\, D( \sigma ||  \rho)
\label{measure}
	\label{6a}
\end{equation}  
where $D$ is any measure of {\em distance} between the two density matrices 
$ \rho$ and $ \sigma$ such that $E({\sigma})$ satisfies the above 
three conditions. To satisfy condition 1) it is sufficient to demand that   
$D( \sigma ||  \rho)=0$ iff 
$\sigma = \rho$. Due to the invariance of ${\cal D}$ under local unitary 
transformations condition 2) is automatically satisfied. For condition 3) 
to be satisfied it is sufficient to demand that $D( \sigma ||  \rho)$ has 
the property that it 
is nonincreasing under every completely positive trace preserving map 
$\Theta$, i.e.
$D(\Theta \sigma || \Theta \rho)\le D( \sigma ||  \rho)$. This can
easily be seen from the following. If $\rho^{*}$ is a separable density 
operator that realizes the minimum of eq. (\ref{6a}) then because 
$\Theta {\cal D} \subset {\cal D}$
we find
\begin{displaymath}
	E({\sigma}) := D({\sigma}||{\rho}^{*}) \nonumber\\
	\ge D(\Theta{\sigma}||\Theta{\rho}^{*}) \nonumber\\
	\ge \min_{{\rho}\in \cal D} D(\Theta{\sigma}||{\rho})
	= E(\Theta  \sigma)\;\; .
\end{displaymath}
The amount of
entanglement given by eq. (\ref{measure}) can be interpreted as finding a 
state $\rho^{*}$ in $\cal D$ that is closest to $\sigma$ under the measure 
$D$. Such a closest state $\rho^{*}$ approximates the classical correlations of 
the state $ \sigma$ `as 
close as possible'. Therefore $E(\sigma)$ measures the remaining quantum 
mechanical correlations. This suggests a division of correlations of the
state $ \sigma$ into two distinct contributions: {\em quantum correlations}, 
$E( \sigma)$, and {\em classical correlations}, 
$D( \rho^* ||  \rho^*_A \otimes  \rho^*_B)$, where $ \rho^*$ is 
the disentangled state that minimizes D and $ \rho^*_A$ and $ \rho^*_B$ 
are its reduced parts (see Fig. 1 for pictorial representation). 

In the following we make special choices for $D( \sigma ||  \rho)$.
First we use an entropic measure of distance between the two density matrices, 
$ \sigma$ and $ \rho$, also called the von Neumann relative entropy, 
which is defined by analogy with the classical Kullback--Leibler distance 
as \cite{Vedral2,Lindblad1,Cover,Lindblad2}:
\begin{equation}
	S({\sigma}||{\rho}) :=  \mbox{tr}\, \left\{ {\sigma} 
	 \ln  \frac{{\sigma}}{{\rho}} \right\}
\end{equation}
where $\ln \frac{{\sigma}}{{\rho}} := \ln{\sigma} - \ln{\rho}$. 
Note that this quantity although frequently referred to as a distance, 
does not
actually satisfy the usual metric properties, e.g. 
$S({\sigma}||{\rho}) \ne S({\rho}||{\sigma})$.
We now define the entanglement of a state ${\sigma}$ to be
\begin{equation}
	E({\sigma}) = \min_{{\rho}\in \cal D}
 	S({\sigma}||{\rho})\;\; .
	\label{14}
\end{equation}
Note that this is a direct generalization of the von Neumann mutual information
which is obtained for $\rho = \sigma_A\otimes \sigma_B$. 
It is now quite easy to check that this measure in fact satisfies conditions 
1)-3),
because it is known that for the relative entropy $S({\sigma}||{\rho})=0$
iff ${\sigma}={\rho}$ and that for any completely positive trace preserving 
map $\Theta$ we have $S(\Theta{\sigma}||\Theta{\rho})  \le S({\sigma}||{\rho})$ 
\cite{Lindblad1,Wehrl1}.

To illustrate some properties of this measure we now restrict ourselves to 
two spin 1/2 subsystems only. First we calculate
$E( \sigma)$ for a pure maximally entangled state.\\
{\bf Proposition 1:}
Entropic entanglement reduces to the von Neumann entropy (of $\ln 2$) for
pure, maximally entangled states defined by 
$|\Phi^{\pm}\rangle = (|00\rangle \pm |11\rangle )/\sqrt{2}$ and 
$|\Psi^{\pm}\rangle = (|10\rangle \pm |01\rangle )/\sqrt{2}$.\\
\noindent
{\em Proof:}
We prove Proposition 1 for the Bell state 
$\sigma\equiv |\Phi^{+}\rangle\langle \Phi^{+} |$. All other maximally 
entangled states can be generated from this one by local unitary transformations
which do not change $E( \sigma)$. As $\sigma$ is a pure state we have
\begin{equation}
	E(\sigma) = \min_{ \rho\in\cal D} tr\{  \sigma \ln 
	\frac{ \sigma}{ \rho} \}
	= \min_{ \rho\in\cal D} - tr\{  \sigma \ln  \rho \}\;\; .
\end{equation} 
Now we use the fact that the function $f(x) = -\ln{x}$ is convex which results
in 
\begin{equation}
	f(\langle \phi| A|\phi\rangle) \le \langle \phi| f( A) |\phi\rangle
\end{equation}
for any operator $A$ and any normalized state $|\phi\rangle$. This leads to
\begin{eqnarray}
	E( \sigma)= 
	\min_{ \rho\in\cal D} -  \langle \Phi^{+}| \ln \rho |\Phi^{+}\rangle
	\ge \min_{ \rho\in\cal D} - \ln \langle \Phi^{+}| \rho |\Phi^{+}\rangle
\end{eqnarray}
It is known \cite{Horodecki2} that 
$ \rho\in{\cal D} \Rightarrow \langle \Phi^{+}| \rho |\Phi^{+}\rangle\le 
\frac{1}{2}$
and therefore $E( \sigma) \ge \ln 2$ . This lower limit can be reached for
example by the state 
$\rho = \frac{1}{2} \left\{ |00\rangle\langle 00| + |11\rangle\langle 11| 
\right\}$.
Therefore we have $E( \sigma)=\ln 2$. ${}_{\Box}$

For any pure, entangled state with coefficients $\alpha$ and $\beta$ (e.g.
$\alpha |00\rangle + \beta |11\rangle$) we conjecture that this measure reduces 
to the usual von Neumann reduced entropy $-|\alpha|^2 \ln |\alpha|^2 - |\beta|^2
\ln |\beta|^2$, but the rigorous proof has not been found.   

Now we also calculate the entanglement of Bell--diagonal states \cite{Horodecki1}. 
We define the density operators 
${\sigma}_{1/2}= |e_{1/2}\rangle\langle e_{1/2}| = |\Psi^{\pm}\rangle\langle
\Psi^{\pm}|$ 
and ${\sigma}_{3/4}= |e_{3/4}\rangle\langle e_{3/4}| =
|\Phi^{\pm}\rangle\langle\Phi^{\pm}|$ where $|\Psi^{\pm}\rangle\, , |\Phi^{\pm}
\rangle$ 
is the usual Bell basis. Then a Bell--diagonal state
has the for $ W = \sum_i \lambda_i {\sigma_{i}}$
We now prove the following\\
{\bf Proposition 2}.
	For a Bell diagonal state ${\sigma}=\sum_{i} \lambda_i {\sigma}_i$ 
	where all $\lambda_i\in [0,\frac{1}{2}]$ we find
	\begin{equation}
		E({\sigma}) = 0 
	\end{equation}
	while for $\lambda_1\ge \frac{1}{2}$ we obtain 
	\begin{equation}
		E({\sigma}) = \lambda_1\ln\lambda_1 + 
	                       (1-\lambda_1) \ln (1-\lambda_1) + \ln 2
	\label{28}
	\end{equation}
 	and analogously for $\lambda_i\ge\frac{1}{2}$.
\noindent

{\em Proof:}
The first case is simple once we
remember that a Bell diagonal state $ \rho$ is separable, i.e. $ \rho\in\cal D$,
iff its spectrum lies in $[0,\frac{1}{2}]$ \cite{Horodecki2}. Therefore 
$E(\sigma)=0$.

To prove the theorem for $\lambda_1\ge \frac{1}{2}$ we again utilize the fact 
that $f(x) = -\ln{x}$ is convex. We obtain
\begin{eqnarray}
	E( \sigma) &=& \sum_{i} \lambda_i \ln{\lambda_i} + 
		      \min_{ \rho\in\cal D} -tr \{ \sigma\ln{\rho} \} 
	\nonumber\\
	         &\ge& \sum_{i} \lambda_i \ln{\lambda_i} + \min_{ \rho\in\cal D} 
	-\sum_{i} \lambda_i \ln{\langle e_i | \rho | e_i\rangle }\;\; .
	\label{30}
\end{eqnarray}
We know that $ \rho\in\cal D$ implies that all $\rho_{ii}\le \frac{1}{2}$ 
(or otherwise the state can be purified \cite{Deutsch,Horodecki1}).
Therefore we can determine the minimum, not over the states from $\cal D$, but 
over the space $\cal B$ of all Bell diagonal states with spectrum in 
$[0,\frac{1}{2}]$. This gives a lower bound to eq. (\ref{30}) because
\begin{displaymath}
	\min_{ \rho\in\cal D} 
	-\sum_{i} \lambda_i \ln{\langle e_i |  \rho | e_i\rangle }
	= \min_{  \rho \in {\cal B} }
	-\sum_{i} \lambda_i \ln{\langle e_i |  \rho | e_i\rangle } \;\; .
\end{displaymath}
Defining $p_i=\langle e_i |  \rho | e_i\rangle$ we have to minimize the function
$f(p_1,p_2,p_3,p_4) = -\sum_{i} \lambda_i \ln{p_i}$ under the constraints
$\sum_{i=1}^{4} p_i = 1$ and $p_i \in [0,\frac{1}{2}]$.
This minimization yields
\begin{equation}
	p_1=1/2 \hspace*{1.cm} p_i = 
	\lambda_i/2(1-\lambda_1)\;\; .
	\label{35}
\end{equation}
The state $ \rho = \sum_{i} p_i\sigma_i$ with the values from eq. (\ref{35}) 
lies in $\cal D$ \cite{Horodecki2} and therefore the lower limit can be reached 
which proves eq. (\ref{28}). ${}_{\Box}$ 

Note that the expression for the entanglement eq. (\ref{28}) given in 
Proposition 2. is different from the entanglement of creation \cite{Bennett1}.
For a Werner state with $F=0.625$ we obtain $\approx 0.04 \ln 2$ whereas the
entanglement of creation is $\approx 0.117 \ln 2$. It is not clear yet what 
these numbers actually mean, and whether they give a bound to the maximum 
possible efficiency of purification schemes. For consistency it is only 
important that if $ \sigma_1$ is more entangled then $ \sigma_2$ for one 
measure than it also must be for all the other measures. Comparing Bennett 
et al's entanglement of creation with our entanglement measure for Bell 
diagonal states shows that this is in fact the case. 

So far we have discussed only the von Neumann relative entropy. However, there
are many other possible distances that we can choose for $D({\sigma}||{\rho})$
in eq. (\ref{6a}) to quantify entanglement of two arbitrarily dimensional 
subsystems. An example of interest is the Bures metric $D_B({\sigma}||{\rho}) 
= 2- 2\sqrt{F(\sigma,\rho)}$, where $F(\sigma,\rho):= \left[tr\{\sqrt{{\rho}}
{\sigma} \sqrt{{\rho}}\}^{1/2}\right]^2 $ is the so called fidelity (or 
Uhlmann's transition probability) \cite{Josza1}. It can be shown that if we
use this distance in eq. (\ref{6a}) we obtain a measure of entanglement that
satisfies the conditions 1)-3) (see \cite{Barum} for the proof that 
fidelity does not decrease under LGM+CC). Other possible measures 
can be found and will be discussed elsewhere. The Bures metric has a 
very nice statistical, operational basis for the measure 
of entanglement in terms of general measurements \cite{Fuchs}. It derives from 
the nature of fidelity as a `measure' of distinguishability between two 
probability distributions $p_{1i} = tr( \sigma  A^{\dagger}_i  A_i) $ and 
$p_{1i} = tr(\rho  A^{\dagger}_i  A_i) $, where 
$\sum_i  A^{\dagger}_i  A_i = {\bf I} $. More precisely,
\begin{equation}
F( \sigma,  \rho) = \min_{ A^{\dagger}_i  A_i} \sum_i 
\sqrt{tr( \sigma  A^{\dagger}_i  A_i)}\sqrt{tr(\rho  A^{\dagger}_i  A_i)}
\end{equation} 
where the minimum is taken over all possible general measurements. This 
possibly enables us in principle to determine eq. (\ref{6a}) and therefore 
also the degree of entanglement experimentally.

So far we have only defined entanglement between two subsystems of arbitrary 
dimensions. It is, 
however, straightforward to generalize this notion to more than two 
subsystems. Let us for simplicity assume that we have three systems, 
$A$, $B$ and $C$. Then the entanglement would be a minimum distance 
of eq. (\ref{6a}) over all disentangled states which in this case would 
be of the form 

\begin{equation}
	 \rho_{ABC} = \sum_i p_i  \rho_{AB} \rho_C + q_i 
	 \rho_{AC} \rho_B + r_i  \rho_A  \rho_{BC}\;\; .
\end{equation}
Again we can see that this class of measures has to satisfy the three imposed
conditions. In the same 
fashion the above approach to quantifying the entanglement could be 
generalized to any number of quantum subsystems. However, the 
complexity involved in minimizing the distance increases with increasing 
the number of the subsystems under consideration.

In this letter we have presented conditions every measure of entanglement 
has to satisfy and shown that there is a {\em whole class} of distance measures
suitable for entanglement measures. The central idea of our construction
is that we calculate the distance between a given state and all possible
disentangled states, taking the minimum as the actual amount of 
entanglement. This construction approximates classical correlations as 
closely as possible and therefore measures the quantum correlations only.
The generalization to entanglement measures for more than two particles
is straightforward.
Our work suggests further investigation is worthwhile into the relationship 
between purification procedures and 
various measures of entanglement suggested above, as well as finding 
a closed form for the expression for entanglement.

We thank the Oxford quantum information group for useful discussions.
This work was supported by the European Community , the UK Engineering and 
Physical Sciences Research Council, by the Alexander 
von Humboldt foundation and by the Knight Trust.

\end{document}